\documentclass[prb,twocolumn,showpacs,superscriptaddress,amsmath,amssymb]{revtex4}

\usepackage{graphicx}
\usepackage{dcolumn}
\usepackage{bm}
\newcommand{\qvec}{{\bf q}}

\newcommand{\kvec}{{\bf k}}
\newcommand{\pvec}{{\bf p}}

\begin{document}
\title{Influence of correlations on transitive electron-phonon couplings 
in cuprate superconductors}
\author{G. Seibold}
\affiliation{Institut f\"ur Physik, BTU Cottbus, PBox 101344, 03013 Cottbus, 
Germany}
\author{M. Grilli}
\affiliation{
Dipartimento di Fisica, Universit\`a di Roma ``La Sapienza'', P.le Aldo
Moro 5, I-00185 Roma, Italy}
\author{J. Lorenzana}
\affiliation{ISC-CNR and
Dipartimento di Fisica, Universit\`a di Roma ``La Sapienza'', P.le Aldo
Moro 5, I-00185 Roma, Italy}
\date{\today}

\begin{abstract}
We investigate a model for the CuO$_2$ plane of high-T$_c$ superconductors
where the charge carriers are coupled 
to A$_{1g}$ and B$_{1g}$ symmetric out-of plane vibrations of the 
oxygen atoms in the presence
of local Hubbard correlations. The coupling is implemented
via a modulation of the hopping integral and we calculate the
renormalization of vertex and pairing scattering functions 
based on the time-dependent Gutziller approximation.
Contrary to local electron-phonon couplings we find that the 
transitive coupling can even be enhanced by correlations for
certain momenta and symmetries of the vibrations. While this effect
may be important for certain properties, we find that, with regard to 
superconductivity, electron-electron  correlations 
still generically lead to a suppression of the pairing
correlations.
Our results allow for an estimate of correlation effects
on the electron-phonon induced pair scattering from weak electron-electron
interactions up to the Mott regime. For onsite repulsions relevant to cuprate
superconductors our calculations reveal a significant contribution
of B$_{1g}$ phonons to d-wave superconductivity. 
\end{abstract}
\maketitle

\section{Introduction}
The search for the new superconducting cuprates by Bednorz and M\"uller
\cite{bm86} was inspired by the idea that in systems containing Jahn-Teller (JT)
ions the strong interaction of electrons with local distortions may lead
to high transition temperatures. While meanwhile the contribution 
of a JT mechanism to 
superconductivity is controversial, the coupling of electrons
to the lattice in cuprates is evidenced in numerous experiments. These
include among others the observation of an isotope effect in scanning
tunneling microscopy (STM) data \cite{lee06}
and in T$_c$ away from optimal doping,~\cite{frank} the softening of
certain phonons under doping,~\cite{queeney99,pintch99,Reich96,Rez06}
and the existence of a critical CuO$_4$ octahedra tilt angle for the
existence of superconductivity  in lanthanum cuprates.~\cite{buech}
Also the kink in the electronic dispersion as seen in angle-resolved
photoemission spectroscopy (ARPES) experiments
has been attributed to strong electron-phonon (el-ph) 
interactions \cite{lanzara01}
which in Bi$_2$Sr$_2$CaCu$_2$O$_{8+\delta}$ compounds is
supported by the observation of an isotope effect in the corresponding
energy scale \cite{iwasawa08} (cf. also Ref. \onlinecite{lan10}
and references therein).
However, density functional theory (DFT) usually yields an
 electron-phonon coupling
constant which is too small in order to explain the high transition  
temperatures of the cuprate materials.~\cite{savra96}
This becomes even worse when the effect of electronic correlations is
considered which are underestimated within the DFT method.
Basically, correlations suppress the charge fluctuations and thus
the scattering of electrons from phonons, which leads to a further reduction
of the el-ph interaction (for a review cf. Refs. 
\onlinecite{gunnar08,marco10}). 

In terms of Hubbard type models the el-ph interaction is usually investigated
for couplings which arise from an expansion of on-site energies
in terms of some vibrational coordinates \cite{kulic93,koch04,gunnar08,entel99}
so that in general both interactions, electronic 
and electron-phonon, are local  or
at least have a dominant local contribution.
 
The interplay of el-el and el-ph interactions leads
to a screening of the latter over the whole Brillouin zone which
is most pronounced at large momenta.~\cite{kulic93,diciolo09}
For sizeable correlations the maximum in the phonon self-energy is then 
shifted from the Peierls wave-vector to $q=0$, which for strong
electron lattice couplings eventually induces a phase separation instability.

Within a slave-boson technique the influence of correlations onto 
the el-ph interaction has also been investigated within a three-band
model where the phonons have been coupled locally to the copper and
oxygen density.~\cite{grilli94} It was found that generically the 
renormalization of the el-ph vertex depends on $v_Fq/\omega$
($v_F$ is the Fermi velocity).  
 It turns out to be strongly suppressed if
this ratio is $\gg 1$, i.e. in the static limit, while it is almost
unaffected in the dynamical limit.

While the interplay between correlations and local (Holstein-type) el-ph
interaction is quite well understood, the situation for 
transitive couplings is less
clear. In a lattice model this type of coupling arises from the modulation
of hopping integrals due to lattice vibrations. Vertex renormalization
effects have been studied
in the context of a tJ-model in Ref. \onlinecite{ishihara04}  and
for the $U\to \infty$ Hubbard model within a slave-boson
scheme in Refs. \onlinecite{kim93,entel98}.
We are not aware of any similar investigation for transitive couplings
within an approach, which can cover the parameter space from weak coupling
up to the Mott regime, which is the task we want to accomplish 
in the present paper.

In cuprates, transitive electron-phonon couplings are believed
to be of significance for the so-called breathing and buckling modes which
correspond to in- and out-of plane motions of the oxygen atoms
and which contribute most to the pairing interaction arising from 
phonons.~\cite{manske,manske2}
For example, the inclusion of a screened Coulomb interaction within the LDA-U
method \cite{zhang07} greatly enhances the el-ph interaction of the
half-breathing mode whereas the influence on the full-breathing mode
is much weaker. An extensive analysis of the strength and material
dependencies of theses phonon modes, including also apical oxygen modes,
has recently been performed in Ref. \onlinecite{johnston}. It was found
that the planar oxygen B$_{1g}$ c-axis modes lead to the strongest
enhancement of d-wave superconductivity. Moreover, 
upon including long-range Coulomb interactions,
which in a layered system lead to low energy plasmon excitations $\Omega_{pl}$,
the coupling can even be overscreened, i.e. the B$_{1g}$ vertex increases
with increasing $\Omega_{pl}$. 
Again the effect of strong local correlations on the coupling of charge carriers
to B$_{1g}$ modes has not been considered in Ref. \onlinecite{johnston}
and is the purpose of the present paper.

Within a one-dimensional model we have previously studied the influence of 
a transitive el-ph interaction on the phonon self-energy 
$\Sigma_{ph}(\qvec,\omega)$.~\cite{oelsen09} 
Most interestingly it was found that $\Sigma_{ph}(\qvec,\omega)$ can be
enhanced for small momenta upon switching on the Hubbard interaction
whereas for large momentum $\Sigma_{ph}(\qvec,\omega)$ becomes suppressed
with $U$ similar to the case of a local (Holstein) el-ph interaction.

\begin{figure}[thb]
\includegraphics[width=7cm,clip=true]{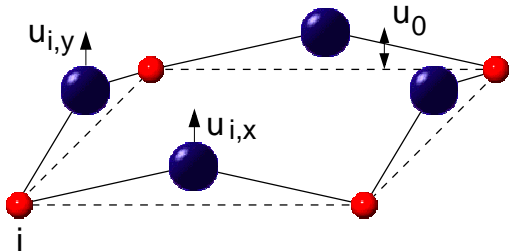}
\caption{(Color online) 
Geometry of the buckled CuO$_2$ plane with Cu (small, red) and
oxygen (large, blue) ions. Electronic degrees of freedom are
on the Cu sites and vibrational degrees on the oxygen sites. Notice that the
configuration where oxygens on the x-bonds are distorted in the negative
z-direction leads to the same model hamiltonian as Eq. (\ref{eq:ham}).}
\label{fig1}
\end{figure}

Motivated by this finding we investigate in the present paper the interplay
between electronic correlations and a transitive el-ph interaction 
with regard to 
the Cooper scattering amplitude.
We exemplify this interplay for the buckling mode where electrons couple 
to out-of-plane (z-direction) 
motions of the oxygen atoms. For a planar CuO$_2$ system, due to symmetry,  
this kind of coupling would be quadratic in the z-coordinates.~\cite{annett95}
However, for vibrations around static $z_0$-displacements of the oxygen atoms
one recovers a linear el-ph interaction. 

The static configuration of a 
frozen-in ${\bf Q}=(0,0)$ mode where all oxygen atoms 
are displaced along the same direction (cf. Fig. \ref{fig1}) 
corresponds to the buckling of the plane in YBa$_2$Cu$_3$O$_7$ (YBCO) 
and Bi$_2$Sr$_2$CaCu$_2$O$_8$ (Bi2212) whereas
the static ${\bf Q}=(\pi,\pi)$ deformation has its realization in the
tilting of the CuO$_4$ octahedra in the low-temperature orthorhombic phase of
La$_{2-x}$Sr$_x$CuO$_4$ (LSCO). Fig.~\ref{fig1} shows the direction 
of positive oxygen displacements ${\bf u}_{i,x(y)}$ on top of the static
deformation. The model which we introduce in the nect section is valid for 
both cases, ${\bf Q}=(0,0)$  and ${\bf Q}=(\pi,\pi)$. However, for the
latter distortion one has to take into account 
a staggered positive direction of ${\bf u}_{ix(y)}$. 
Therefore  our investigations hold for a broad class of
cuprate materials. 

For the sake of clarity it should be stated that the interplay between 
correlations and el-ph interactions strongly depends on the 
model under consideration.
For example, a coupling to local density fluctuations in the three-band
model for cuprate superconductors may transform into a transitive
coupling when one derives
an effective low-energy one-band model by eliminating the high energy
degrees of freedom (cf. e.g. Ref. \onlinecite{dev04}). 
In our paper we are concerned with a one-band hamiltonian with transitive
electron-phonon interactions which can be contrasted with analogous
calculations for a local Holstein-type coupling in Ref.~\onlinecite{diciolo09}.
Nevertheless one should keep in mind that our results may
also mimic the local el-ph coupling in related multiband models.

Our paper is organized as follows. In Sec. \ref{sec2} we outline the
formalism of our calculation which is based on the time-dependent
Gutzwiller approximation. Sec. \ref{sec3} presents the results of 
the renormalization
of the transitive EPI due to correlation effects and we conclude
our investigations in Sec. \ref{sec4}. In both Sec. \ref{sec2} and
Sec. \ref{sec3} we additionally discuss a one-dimensional model
for illustrative purposes.

\section{Formalism}\label{sec2}
We use an effective one-band hamiltonian where the hopping between 
neighboring sites  is modulated by the displacement
of adjacent oxygen atoms from their distorted equilibrium 
position.~\cite{note1}
The geometry of the distorted Cu$O_2$ plane is sketched in Fig. \ref{fig1}. 
Similar models have been investigated in Refs. \onlinecite{naz96,norm96}.

The hamiltonian reads   
\begin{eqnarray}
H&=&\sum_{ij\sigma} t_{ij} c_{i\sigma}^\dagger c_{j\sigma}
+ U \sum n_{i,\uparrow}n_{i,\downarrow}\label{eq:ham} \\
&+& g_{tr}\sum_{i\sigma,\alpha}u_{i,\alpha}(c_{i\sigma}^\dagger c_{i+\alpha,\sigma}
+ h.c.)+\sum_{i,\alpha}\left[\frac{p_{i\alpha}^2}{2M} + \frac{1}{2}
\omega_0^2 u_{i,\alpha}^2\right]\nonumber 
\end{eqnarray}
Here $c_{i\sigma}^{(\dagger)}$ creates (annihilates) an electron on
a square lattice and the vibrational degrees of freedom are sitting
on the bonds. The parameter $t_{ij}$ is the hopping amplitude between
sites $i$ and $j$. $U$ denotes the local Hubbard repulsion and
$u_{i,\alpha}$ with $\alpha=x,y$ is the displacement
from equilibrium of the (oxygen) atom at position $i+\alpha/2$
(cf. Fig. \ref{fig1}). $g_{tr}$  is the coupling parameter between
lattice displacement and electronic nearest-neighbor hopping. 
The last term in Eq. (\ref{eq:ham}) 
corresponds to kinetic and elastic energy of the oxygens.

Our approach is based on the time-dependent Gutzwiller approximation
(TDGA) \cite{sei01} which gives an accurate description of
charge fluctuations of the Hubbard model \cite{sei01,sei03} 
and which previously has been succesfully applied 
to the study of vertex renormalizations in the 
Hubbard-Holstein model.~\cite{diciolo09}
The idea is to evaluate the energy of Eq. (\ref{eq:ham}) in the presence
of an arbitrary lattice distortion and charge fluctuation within the 
Gutzwiller approximation (GA) \cite{gut65,kot86,geb90} and to expand
the functional up to second order in the lattice $Q_q^\pm$ and 
electronic density  fluctuations. 

The Gutzwiller variational wave
function has the form $|\Psi_g \rangle =P_g |SD\rangle$ where  $P_g$ is 
the Gutzwiller
projector and $|SD\rangle$ a Slater determinant. For $|SD\rangle$ we
use a state with an arbitrary charge profile. We define the associated
one-body density as  $\rho_{ij\sigma}=\langle
SD|\hat c_{j\sigma}^\dagger \hat c_{i\sigma}  |SD\rangle$. Since
magnetic and charge excitations decouple we can restrict to Slater
determinants with  $\rho_{ij\uparrow}=\rho_{ij\downarrow}$.

Our
expansion below will be in terms of the density fluctuations of the
SD which physically describes quasiparticles and not physical
electrons.\cite{vol84} The two densities coincide in the 
onsite case $\rho_{ii\sigma}$  while for off-diagonal density the real particle  
densities  are given by $z_{i\sigma}z_{j\sigma}
\rho_{ij\sigma}$ where  $z_{i\sigma}$ denote  the Gutzwiller renormalization 
factors.

The quasiparticle density fluctuations that enter into the problem are
\begin{eqnarray}
\delta {\bf X_q} &=& \sum_{k,\sigma} {\bf V}(\kvec,\qvec)
\delta \rho_{k+q,k,\sigma} \nonumber \\
 &=& \sum_{k,\sigma} \left(\begin{array}{c}
1 \\ \epsilon^0(\kvec)+\epsilon^0(\kvec+\qvec) \\
2 \cos(k_x+\frac{q_x}{2})\\
2 \cos(k_y+\frac{q_y}{2})\\
\end{array}\right)\delta \rho_{k+q,k,\sigma} \label{eq:xv}
\end{eqnarray}
with $\epsilon^0(\kvec)$ being the Fourier transformed of the hopping 
$t_{ij}$. In the vector Eq. (\ref{eq:xv}) the first two components 
$\delta X^1_q$ and $\delta X^2_q$ correspond to local and 
transitive density fluctuations, the latter being generated in the
expansion of the kinetic energy. The $\delta X^3_q$ and
$\delta X^4_q$ components are also transitive fluctuations but arise from the
expansion of the el-ph coupling. 

The expansion results in 
\begin{equation}\label{eq:expand}
\delta E^{GA} = E_{ee}^{(2)}+E_{coup}^{(2)}+E_{ph}^{(2)} .
\end{equation}
where the first term describes the interaction between 
electronic density fluctuations
\begin{equation}\label{eq:eexp}
E_{ee}^{(2)} =\frac{1}{2N}\sum_{q} \left(\begin{array}{c}
\delta X^1_q \\ \delta X^2_{q}
\end{array}\right)
{\underline{W_{q}}}^{ee}
\left(\begin{array}{c}
\delta X^1_{-q} \\  \delta X^2_{-q}
\end{array}\right)
\end{equation}
and the elemens of the interaction kernel
\begin{eqnarray}
{\underline{W_{q}}}^{ee} =
\left(\begin{array}{cc}
A_{q} & B_{q}\\
B_{q} & C_{q}\\
\end{array}\right)
\label{Wqgen}
\end{eqnarray}
are defined in the appendix.

The interaction between electronic density and lattice fluctuations 
$Q_q^{\alpha}=1/\sqrt{N}\sum_i \exp{[-i\qvec R_{i+\alpha/2}]}u_{i,\alpha}$ 
is described by the second term
\begin{equation}\label{eq:coup}
E_{coup}^{(2)}=\frac{1}{\sqrt{N}}\sum_{q,i,\alpha=x,y} \gamma_q^{i,\alpha}
\delta X_q^i Q_{-q}^\alpha
\end{equation}
with
\begin{eqnarray}
\gamma_q^{1,\alpha}&=&2 g_{tr}K B_q\cos(\frac{q_\alpha}{2})\\
\gamma_q^{2,\alpha}&=& 2 g_{tr}K C_q \cos(\frac{q_\alpha}{2}) \\
\gamma_q^{3,\alpha}&=& g_{tr}z_0^2 \delta_{\alpha,x} \label{gacoup}\\
\gamma_q^{4,\alpha}&=& g_{tr}z_0^2 \delta_{\alpha,y}
\end{eqnarray}
and $K\equiv \frac{2}{N}\sum_{k,\sigma}\cos(k_x)\langle n_{k,\sigma}\rangle$.
$z_0$ denotes the hopping renormalization factor in the GA
which generically decreases with $U$ and at half-filling vanishes
at the Brinkmann-Rice transition $U_{BR}$.~\cite{ubr}

In the derivation of Eqs. (\ref{eq:eexp})-(\ref{eq:coup}) [and the
following Eq. (\ref{eq:eph})] we have already 'antiadiabatically' 
eliminated \cite{sei01,sei03,diciolo09,oelsen09} the double
occupancy fluctuations which arise in the expansion of the GA functional
within the TDGA prescription.
For the present model the corresponding formal procedure is analogous to that 
described in Ref. \onlinecite{oelsen09}.

Notice that on the GA level the coupling to the lattice fluctuations
is reduced $\sim z_0^2$ by the correlations. However, the TDGA induces
additional couplings to local ($\delta X_\qvec^1$) and transitive
($\delta X_\qvec^2$) density fluctuations and we will show below that
in certain cases this will 
even lead to an enhancement of the electron-phonon vertex.

Finally we obtain for the lattice energy
\begin{equation}\label{eq:eph}
E_{ph}^{(2)}=\frac{1}{2N}\sum_{q,\alpha\beta} 
M (\Omega^{\alpha\beta}_q)^2 Q^\alpha_q Q^{\beta}_{-q} 
\end{equation}
and it turns out that the elimination of the double occupancy introduces a  
novel renormalization and a mixing of the $x,y$-buckling modes:     
\begin{eqnarray}\label{eq:omeff}                                    
(\Omega^{\alpha\beta}_q)^2&=&\omega_0^2 \delta_{\alpha\beta}        
-\kappa_q^2 \cos(\frac{q_\alpha}{2})\cos(\frac{q_\beta}{2})\\       
\kappa_q^2&=&-\frac{1}{M}(2 g_{tr})^2 K_\alpha K_\beta C_q .      
\end{eqnarray}

The eigenmodes are obtained from the transformation
\begin{equation}\label{eq:trafo}                                   
\left(\begin{array}{c}                             
Q_q^x \\ Q_q^y                                     
\end{array}\right)                                 
= {\underline{U}}                                                 
\left(\begin{array}{c}                             
V_q^{B1g} \\ V_q^{A1g}                                     
\end{array}\right)                                 
\end{equation}
with
\begin{displaymath}             
{\underline{U}}
= \frac{1}{\sqrt{\cos^2(\frac{q_x}{2})+\cos^2(\frac{q_y}{2})}}
\left(\begin{array}{cc}                             
\cos(\frac{q_y}{2}) & \cos(\frac{q_x}{2}) \\
-\cos(\frac{q_x}{2}) & \cos(\frac{q_y}{2})
\end{array}\right)
\end{displaymath}
and correspond to out- (i.e. $B_{1g}$ symmetry) and in- (i.e.
A$_{1g}$ symmetry) phase oscillations of the $x/y$ oxygen atoms.

The eigenfrequencies read as
\begin{eqnarray}
\omega_q^{B_{1g}}&=&\omega_0 \nonumber \\
\omega_q^{A_{1g}}&=&\sqrt{\omega_0^2-\kappa_q^2 (\cos^2(\frac{q_x}{2})
+\cos^2(\frac{q_y}{2}))}. \nonumber 
\end{eqnarray}

Thus the frequency of the $A_{1g}$ mode softens due to  
the elimination of the double occupancy fluctuations
as a natural consequence of the interaction between
both bosonic degrees of freedom. There will be an additional softening
due to phonon self-energy effects. 
In the following we plot results in dependence
of the coupling parameter $\lambda=g_{tr}^2/\omega_0$ using
the {\it bare} phonon frequencies. One should therefore keep in mind
that the inclusion of the correlation
induced phonon softening would give an additional
(small) enhancement in the A$_{1g}$ channel. 

In order to compute vertex corrections and the effective interaction
between quasiparticles we introduce correlation functions $\chi^{ij}_\qvec$
which describe the response of 
the quasiparticle  (Slater determinant) density fluctuations $\delta X_q^i$ 
[Eq. (\ref{eq:xv})]
to an external perturbation $\delta \lambda_\qvec^i$ via
$$X_{\bf q}^i= \chi^{ij}_\qvec \lambda_\qvec^j.$$

It is convenient to represent the non-interacting 
susceptibilities as 
\begin{displaymath}
\underline{\underline{\chi^{0}_{q}(\omega)}}=
\left(\begin{array}{cc}
{\underline{\chi^{ee,0}_{q}}} & {\underline{\chi^{ex,0}_{q}}} \\
{\underline{\chi^{xe,0}_{q}}} & {\underline{\chi^{xx,0}_{q}}} \end{array}\right)
\end{displaymath}
where e.g. the $2\times 2$ susceptibility
matrix ${\underline{\chi^{ee,0}_{q}}}$ corresponds to
the submatrix of $\chi_{ij}(\qvec)$ for $i,j=1,2$ and the definition
of the other block matrices follows analogously. Notice that here and in
the following we denote $4\times 4$ ($2\times 2$) matrices with
a double (single) underscore.
 
The lattice induced (bare) interaction is obtained by assuming 
that the phonon frequencies are larger than the frequencies 
of the electronic excitations of interest.
 Therefore we can eliminate the lattice eigenmodes for a 
frozen electronic configuration, 
$\partial E^{GA}/\partial V_q^s = 0$ ($s=B_{1g}$, $A_{1g}$). 
One obtains 
in addition to Eq. (\ref{eq:eexp}) a lattice induced (bare) interaction 
between density fluctuations
\begin{equation}\label{velph}
V^{ee}_{ij}(\qvec)= -\frac{1}{2N} \sum_{q,s,\alpha\beta}
\frac{1}{M\omega_q^{s}} \gamma_q^{i,\alpha}\gamma_{-q}^{j,\beta}
U_q^{\alpha,s}U_{-q}^{\beta,s} .
\end{equation}
Using the TDGA, this interaction together with the bare Hubbard
part Eq. (\ref{eq:eexp}) can be dressed by the electronic
density fluctuations similar to conventional RPA theory.
We define the 'undressed' interaction from the sum of Eqs. 
(\ref{Wqgen},\ref{velph})
\begin{equation}
\underline{\underline{V^0(\qvec)}}=\underline{\underline{V^{ee}(\qvec)}}
+ \left(\begin{array}{cc}
{\underline{W_{q}^{ee}}} & {\underline{0}} \\
{\underline{0}} & {\underline{0}}\\
\end{array}\right)
\end{equation}
which upon resumming the RPA leads to the effective dressed interaction
between quasiparticles
\begin{equation}                                                             
\underline{\underline{{W^{eff}(\qvec)}}} = 
\underline{\underline{{V^0(\qvec)}}}\left\lbrack {\bf 1}
- \underline{\underline{\chi_q}}^{0}\underline{\underline{{V^0(\qvec)}}}
\right\rbrack^{-1}  \label{veff}.
\end{equation}

Finally we can use Eq. (\ref{veff}) to define the Cooper pair
scattering amplitude
\begin{equation}\label{gamfin}
\Gamma(\kvec,\qvec)=
\sum_{i,j}W^{eff}_{ij}(\qvec)
V_i(\kvec,\qvec) V_j(-\kvec,-\qvec)
\end{equation}
and the $V_i(\kvec,\qvec)$ have been defined in Eq. (\ref{eq:xv}).
We will usually take both $\kvec$ and $\kvec+\qvec$ on the Fermi surface
and restrict to the static limit ($\omega=0$).

We can expand  Eq. (\ref{gamfin}) up to order $\sim g_{tr}^2$ in the
phonon contribution. The coefficient of $\sim g_{tr}^2$ defines the
second order el-ph contribution 
to the pair scattering amplitude as 
\begin{equation}\label{gamfin2nd}
\Gamma^{(2)}_s(\kvec,\qvec)=-\frac{1}{\omega^s_q}
\sum_{i,j} \widetilde{\gamma}_q^{i,s} \widetilde{\gamma}_{-q}^{j,s}
V_i(\kvec,\qvec) V_j(-\kvec,-\qvec) .
\end{equation}
This scattering amplitude does not include the direct scattering
amplitude due to el-el interactions. It includes only el-ph
interactions which are, however, renormalized by the el-el interactions.  

Here the dressed vertex functions are given by
\begin{widetext}
\begin{equation}\label{renv}
\widetilde{\gamma}_q^{i,s}=\left(\begin{array}{cc}
\left\lbrack
{\underline{1}}-
{\underline{W_{q}}}^{e-e}{\underline{\chi^{ee,0}_{q}}}
\right\rbrack^{-1} &
{\underline{W_{q}}}^{e-e}\left\lbrack
{\underline{1}}-
{\underline{\chi^{ee,0}_{q}}}{\underline{W_{q}}}^{e-e}
\right\rbrack^{-1}{\underline{\chi^{ex,0}_{q}}} \\
 & \\
0 & 1
\end{array}\right)_{ij}\gamma_q^{j,\alpha}U_q^{\alpha,s} .
\end{equation}
\end{widetext}

From Eqs. (\ref{renv},\ref{gamfin2nd}) it turns out that the 
original coupling between buckling modes
and density fluctuations ($\gamma_q^{3,x}$, $\gamma_q^{4,y}$), 
Eq. (\ref{gacoup}),
does not get renormalized by the TDGA but it is only reduced via
the $z_0$ factors already present on the GA level. This also
implies that within standard (Hartree-Fock based, i.e. $z_0=1$) perturbation theory
a transitive el-ph interaction remains unrenormalized up
to second order in the coupling constants. 
On the other hand,  the TDGA can even induce an 'antiscreening' of the 
local and transitive fluctuations, leading to an enhancement of the
$\widetilde{\gamma}_q^{i,s}$ as compared to the bare 
couplings ${\gamma_q^{i,\alpha}}$ (cf. next section).

\subsubsection*{Illustrative one-dimensional model}

\begin{figure}[htbp]
\includegraphics[width=8cm,clip=true]{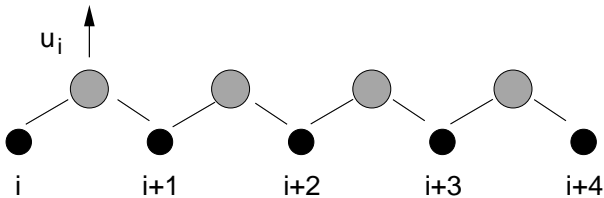}
\caption{One-dimensional CuO model where the oxygens (light shaded circles)
oscillate perpendicular to the chain. The hopping between nearest-neighbor
Cu (full circles) atoms $i$ and $i+1$ is then linearly expanded in $u_i$.}
\label{fig:fig1d}
\end{figure}

In order to better illustrate the influence of
correlations on the el-ph interaction within our TDGA approach, we will consider also
a simpler one-dimensional model (cf. Fig.~\ref{fig:fig1d}).
The result of this and the two-dimensional system will be presented in 
Sec. \ref{sec3}.

The transformation of the previously developed formalism 
to the one-dimensional model is straightforward.
Since there is only one vibrational degree of freedom the relevant fluctuations
are 
\begin{eqnarray}
\delta {X_q} &=& \sum_{k,\sigma} {V}(k,q)
\delta \rho_{k+q,k,\sigma} \nonumber \\
 &=& \sum_{k,\sigma} \left(\begin{array}{c}
1 \\ \epsilon^0(k)+\epsilon^0(k+q) \\
2\cos(k+q/2)
\end{array}\right)\delta \rho_{k+q,k,\sigma}
\label{eq:xv1d}
\end{eqnarray}
with a dispersion $\epsilon^0_k=-2t\cos(k)$.   
For the bare vertices one obtains
\begin{eqnarray}
\gamma_q^{1}&=&2 g_{tr}K B_q \cos(q/2)\label{1dgam1}\\
\gamma_q^{2}&=& 2 g_{tr}K C_q \cos(q/2) \label{1dgam2} \\
\gamma_q^{3}&=& g_{tr}z_0^2 \label{gacoup1d}
\end{eqnarray}
with $K=\frac{4}{\pi} sin(k_F)$. 

\section{Results}\label{sec3}
\subsection{One-dimensional case}
\begin{figure}[htbp]
\includegraphics[width=8.5cm,clip=true]{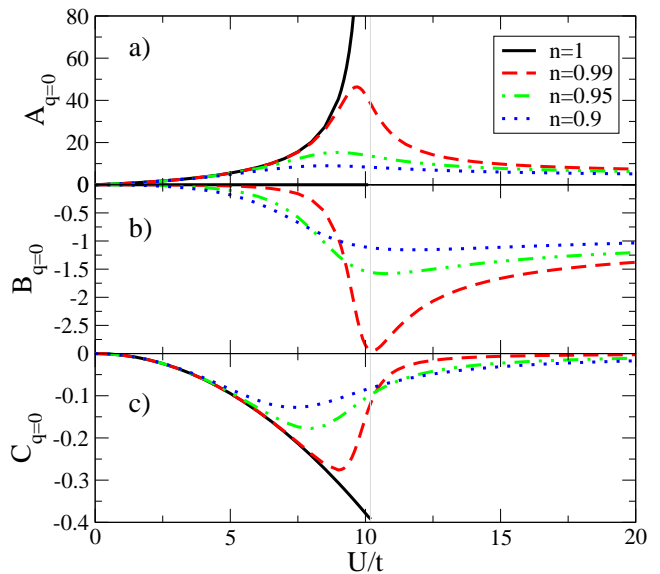}
\caption{(Color online)
(Color online) Elements of the electronic interaction kernel Eq.~(\ref{Wqgen})
for $q=0$ as a function of $U$ and different fillings in the one-dimensional CuO model.
The vertical line at $U/t=32/\pi$ corresponds to the Brinkmann-Rice transition.}
\label{fig:wq1d}
\end{figure}

Before turning to the two-dimensional cuprate model it is instructive
to analyze the model for a linear CuO chain, where the oxygens oscillate
perpendicular to the chain direction. 
Fig. \ref{fig:wq1d} shows the elements of the electronic interaction
kernel Eq.~(\ref{Wqgen}) for $q=0$. $A_{q=0}$ corresponds to the
interaction between local density fluctuations. Close to half-filling it
develops a pronounced maximum at the 
Brinkmann-Rice transition $U_{BR}$
where it actually diverges at exactly $n=1$.
The coefficient $B_{q=0}$ mixes local and transitive density fluctuations
and vanishes at half-filling. However, away from $n=1$, $|B_q|$ increases upon
approaching $U_{BR}$ and also stays finite beyond this value [cf. 
Fig. \ref{fig:wq1d}(b)]. Finally, the coefficient $C_q$ [Fig. \ref{fig:wq1d}(c)]
corresponds to the
interaction between transitive density fluctuations. It also
develops a maximum close to $U_{BR}$, but vanishes in the $U\to \infty$ limit.

\begin{figure}[htbp]
\includegraphics[width=8.5cm,clip=true]{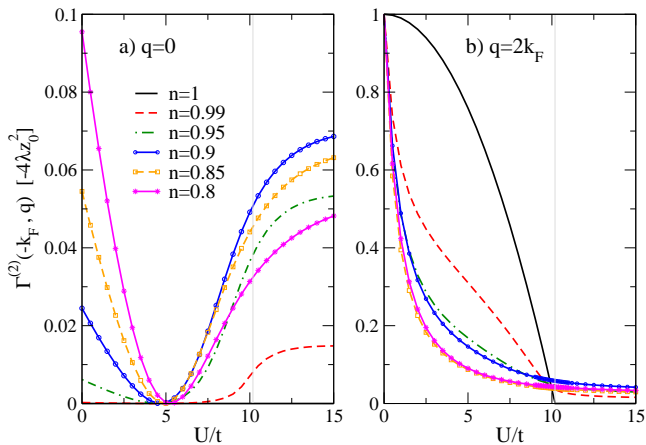}
\caption{(Color online) Second order pair scattering 
amplitude $-\Gamma^{(2)}(\kvec,\qvec)/(4\lambda z_0^2)$ 
for the one-dimensional model as a function of $U$. a) Scattering with $\qvec=0$ 
at $\kvec=-\kvec_F$; b) Scattering between $\pm \kvec_F$ with $\qvec=2\kvec_F$.
The vertical line at $U/t=10.19$ marks the Brinkmann-Rice transition.}
\label{fig:gam1d}
\end{figure}

Let us now turn to the discussion of the renormalized vertex functions via the
scattering amplitude $\Gamma^{(2)}/z_0^2$
which is shown in Fig. \ref{fig:gam1d}. 
Notice that we divide by the GA renormalization factor which
originates from the density of states renormalization  $N^*=N_0/z_0^2$ within
the GA. Indeed, upon plotting $\Gamma^{(2)}/z_0^2$ we take into account the
interaction in both quantities.
We point out that the bare
coupling on the GA level is $\sim g_{tr}z_0^2$, which for the 'bare' scattering
amplitude would result in a reduction of $\Gamma^0/z^2_0 \sim g_{tr}^2 z_0^2$ 
upon increasing $U/t$. 

In the one-dimensional model we have just two Fermi points
so that the scattering of pairs on the 'Fermi surface' is restricted to 
$q=0$ and $q=2k_F$, respectively. 
Consider first the  $q=0$ case which is shown in
Fig. \ref{fig:gam1d}(a).
At half-filling the scattering amplitude is identically zero
since in this case the chemical potential is at $E_F=0$ so that the
only non-vanishing fluctuations are the local ones [cf. Eq. (\ref{eq:xv1d})
where for $\qvec \to 0$ the fluctuations are restricted to $k=k_F=\pm \pi/2$
and $\varepsilon^0(k_F)=0$]. 
Although these are coupled to the vibrations via the coefficient $B_q$ 
[cf. Eq. (\ref{1dgam1})], this latter also vanishes at $n=1$, leading
to the vanishing of the scattering amplitude. 
Slightly below half-filling (dashed line in Fig. \ref{fig:gam1d}a) one
obtains an increase in $\Gamma^{(2)}/z^2_0$ which can be shown to be due to
the behavior of the renormalized coupling 
 $\widetilde{\gamma}_{q=0}^{2}
\sim B_{q=0}/(1+A_{q=0}N^*)$ to the transitive density fluctuations. 
It is the concomitant increase of $|B_{q=0}|$ around $U_{BR}$ together
with the reduced screening from local density fluctuations ($\sim
1/(1+A_{q=0}N^*)$ for $U/t > U_{BR}$, which leads to the peculiar
behavior of  $\Gamma^{(2)}/z^2_0$ slightly below half-filling and $q=0$.
Remarkably, close to half-filling the scattering amplitude is amplified
by large interactions and it becomes many times larger than the 
scattering amplitude in the non interacting case. This
is just the opposite of what happens in the case of Holstein like
couplings. 

For larger values of the doping the bare coupling $\gamma_{q=0}^{3}$
dominates the value of $\Gamma^{(2)}/z^2_0$ for small values of $U/t$ 
which therefore acquires a minimum between $U/t=0$ and $U_{BR}/t$.

In the case of $q=2k_F$ scattering the behavior is simpler.
At half-filling ($k_F=\pi/2$) it can be seen from Eqs. (\ref{1dgam1},
\ref{1dgam2}) that only the bare coupling ($g_{tr}z_0^2$) survives.
For $n=1$ this is not affected by the screening from density fluctuations.
The $n=1$ result in Fig. \ref{fig:gam1d}b (solid line) therefore simply reflects
the behavior of $z_0^2$ as a function of $U/t$ and vanishes  at $U_{BR}=32t/\pi$.
Away from half-filling the additional screening leads to a further
reduction of the scattering amplitude which continuously decreases with
increasing $U/t$ even beyond $U_{BR}$.

\begin{figure}[htbp]
\includegraphics[width=8cm,clip=true]{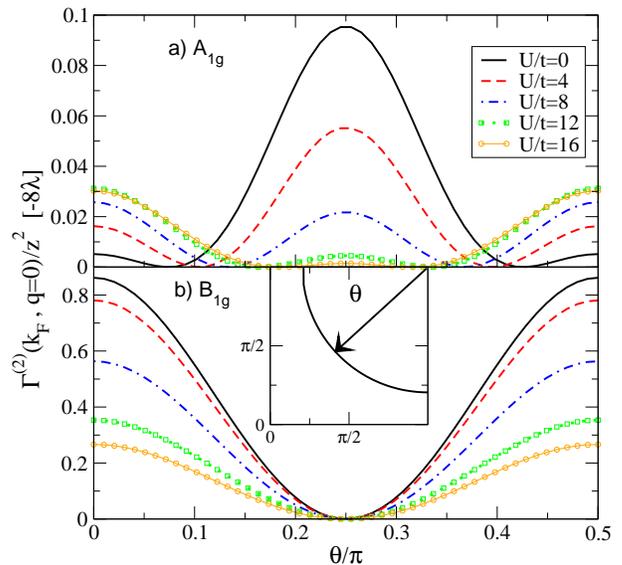}
\caption{(Color online)
Scattering function $-\Gamma^s(\kvec_F,\qvec=0)/z_0^2$ 
(in units of $8\lambda$) for various values of $U/t$ and zero
momentum transfer of pairs on the Fermi surface.
Coupling to A$_{1g}$ (a) and B$_{1g}$ (b) modes.
Parameters: $\delta=0.1$, $t'/t=-0.4$.}
\label{fig:scaq0}
\end{figure}

\subsection{Two-dimensional model}
From the analysis of the 1D model we have learned that within our
TDGA analysis, correlations can enhance the vertex  
for small momentum transfer. On the other hand, correlations
generically reduce the coupling  
for large $q$ scattering. We now investigate whether this result
persists for the model in two dimensions, 
introduced at the beginning of the previous section.

In the following figures we discuss  the renormalized vertex function from
the second order pair scattering amplitude
$\Gamma^{(2)}(\kvec_F,\qvec_F)/z_0^2$ [cf. Eq. (\ref{gamfin2nd})] 
in units of $-8 \lambda$ which
makes this quantity dimensionless and positive. 
The factor of 'eight' originates from
the definition of the bare coupling Eqs. (\ref{eq:coup}, \ref{eq:trafo}) 
which (for $\qvec=0$) is given by $g_{tr}z_0^2 \sqrt{2}
\left[\cos(k_x) - \cos(k_y)\right]$
for B$_{1g}$- and $g_{tr}z_0^2 \sqrt{2}
\left[\cos(k_x) + \cos(k_y)\right]$ 
for A$_{1g}$ symmetry so that $8g_{tr}^2/\omega_0\equiv8\lambda$ 
corresponds to the maximum scattering amplitude for $U/t=0$.

In Fig. \ref{fig:scaq0} we show the second order scattering function 
Eq. (\ref{gamfin2nd}) for pairs on the Fermi surface and zero momentum transfer, i.e. 
$\Gamma^{(2)}(\kvec_F,\qvec=0)$. For  $A_{1g}$-symmetry the coupling function
is $\sim \cos(k_x)+\cos(k_y)$  so that the $U=0$ scattering is largest
around the nodes. For a $B_{1g}$-type coupling  
$\sim \cos(k_x)-\cos(k_y)$ the $U=0$ scattering is maximized around
the antinodes. 
This interaction becomes continuously suppressed upon increasing $U/t$. 
Remarkably the vertex function from the coupling to $A_{1g}$ modes 
is instead enhanced around the antinodal regions while it is also
suppressed around the nodes. As in the one-dimensional case,  
the enhancement by the interaction contrasts the corresponding findings
for local couplings.

\begin{figure}[thbp]
\includegraphics[width=8cm,clip=true]{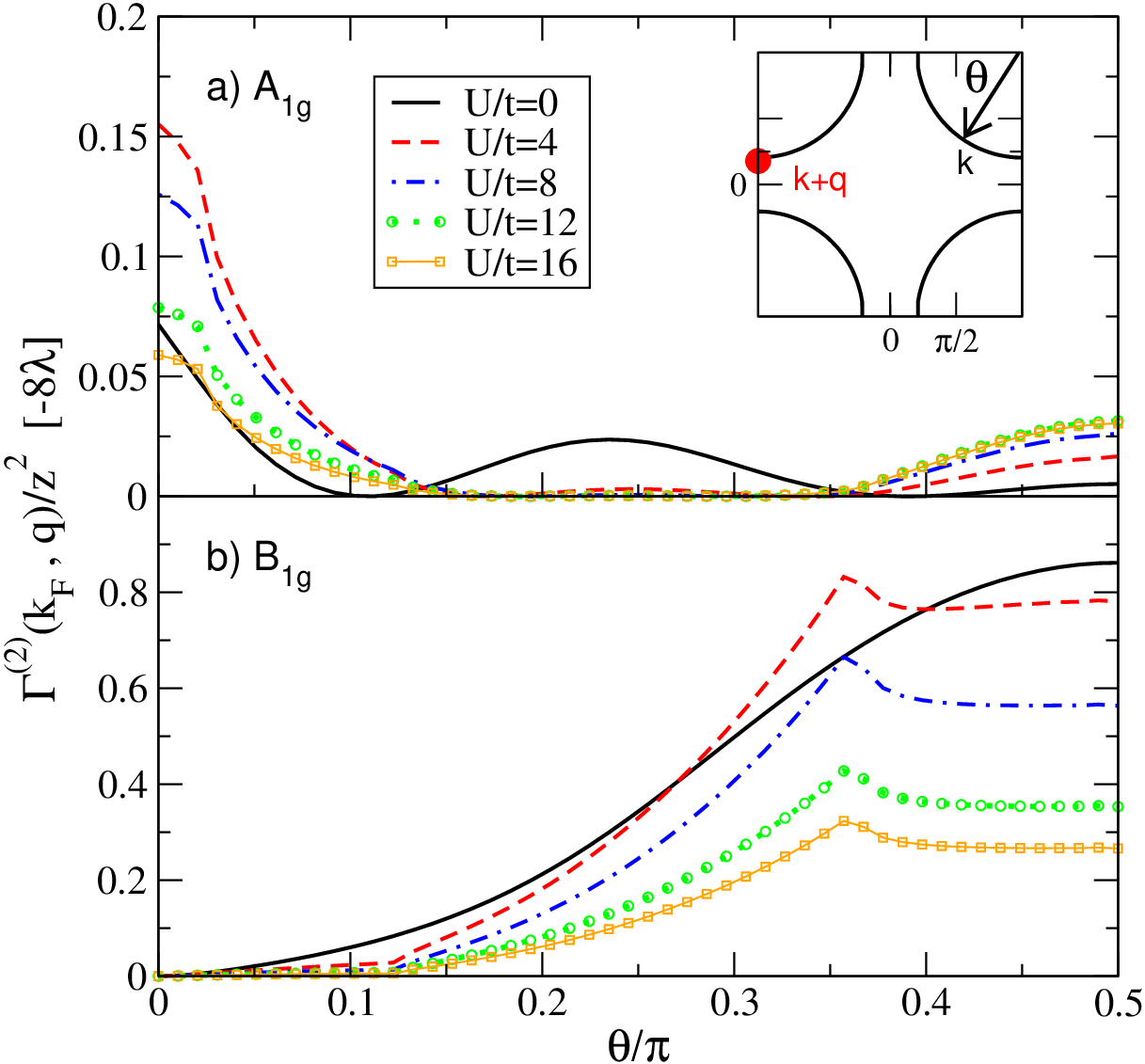}
\caption{(Color online) Scattering function $-\Gamma^{(2)}(\kvec_F,\qvec)$ for  
pairs on the Fermi surface as indicated in the inset.
Coupling to A$_{1g}$ (a) and B$_{1g}$ (b) modes.
Dressed (solid) and bare (dashed) vertex along selected
symmetry directions in the Brillouin zone.  
$\delta=0.1$, $t'/t=-0.4$.}
\label{fig:scaq04}
\end{figure}

\begin{figure}[thbp]
\includegraphics[width=8cm,clip=true]{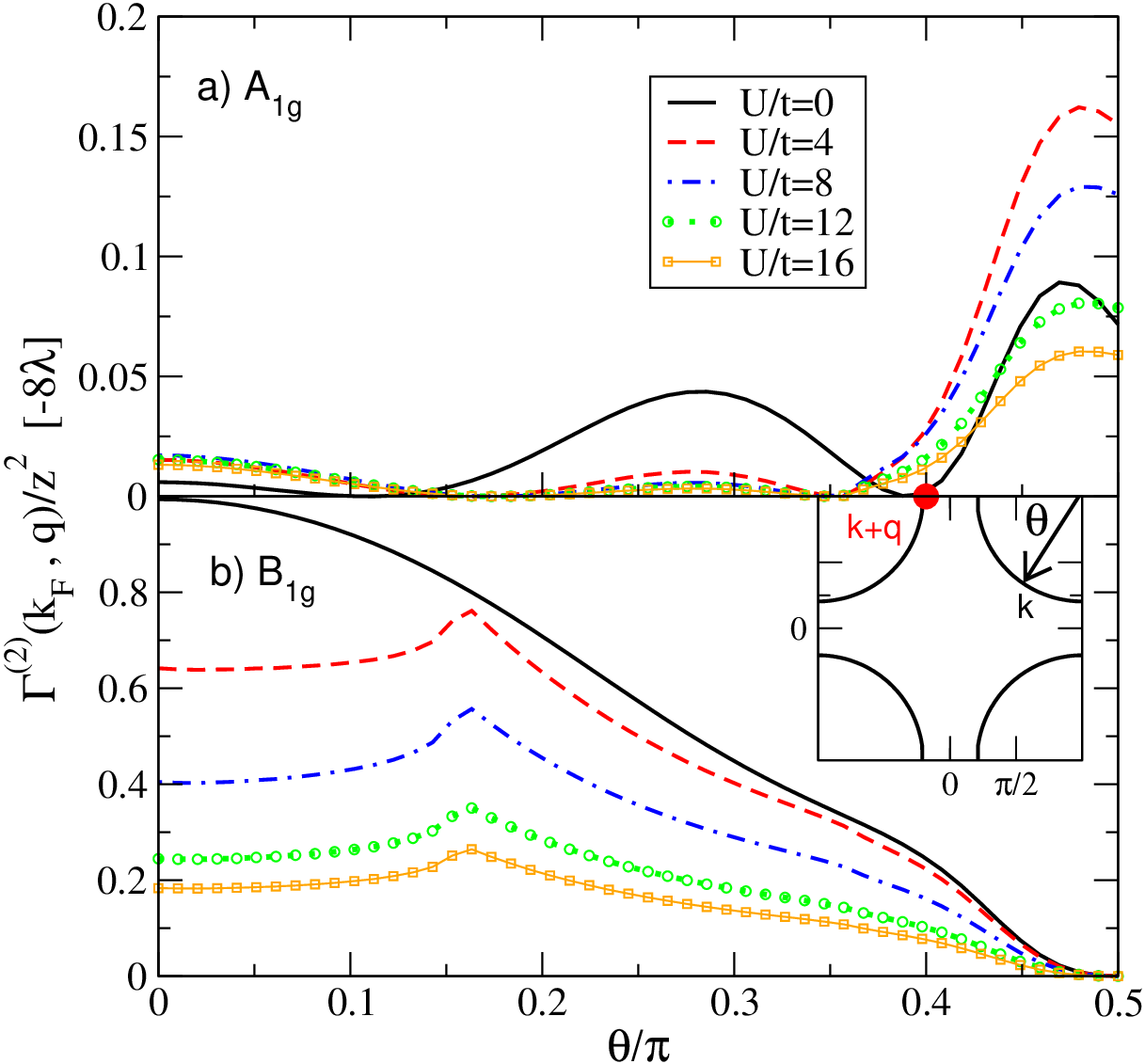}
\caption{(Color online) Scattering function $-\Gamma^s(\kvec_F,\qvec)$ for  
pairs on the Fermi surface as indicated in the inset.
Coupling to A$_{1g}$ (a) and B$_{1g}$ (b) modes.
Dressed (solid) and bare (dashed) vertex along selected
symmetry directions in the Brillouin zone.  
$\delta=0.1$, $t'/t=-0.4$.}
\label{fig:scaq05}
\end{figure}

Figs. \ref{fig:scaq04}, \ref{fig:scaq05}, \ref{fig:scaq06} 
display $\Gamma^{(2)}(\kvec,\qvec)/z_0^2$  for a fixed momentum
$\kvec+\qvec$ of the scattered pair on the FS (cf. inset).

For the scattering from the A$_{1g}$-type mode the generic trend is a
correlation induced enhancement of the vertex function when the
pairs are scattered between the antinodal regions.
In all other cases correlations  generally suppress
the scattering from the A$_{1g}$ mode.
Since the enhancement occurs only for those regions in momentum space
$\kvec \approx \kvec_{antinodes}$ where the vertex is small, the overall
effect is thus a suppression of the coupling upon increasing $U/t$.

In the case of scattering from the B$_{1g}$ mode, 
one observes a suppression for all momenta $\kvec$ and $\kvec+\qvec$.
In fact, as in the one-dimensional case a correlation induced 
enhancement of the vertex is due to the coupling to transitive density
fluctuations (i.e. $\delta X_q^2$). It turns out that this coupling
vanishes for B$_{1g}$ symmetry leading to a generic suppression
of the coupling in contrast to the A$_{1g}$ mode.

\begin{figure}[thbp]
\includegraphics[width=8cm,clip=true]{fig8.eps}
\caption{ (Color online) Scattering function $-\Gamma^s(\kvec_F,\qvec)$ for  
pairs on the Fermi surface as indicated in the inset.
Coupling to A$_{1g}$ (a) and B$_{1g}$ (b) modes.
Dressed (solid) and bare (dashed) vertex along selected
symmetry directions in the Brillouin zone.  
$\delta=0.1$, $t'/t=-0.4$.}
\label{fig:scaq06}
\end{figure}

\begin{figure}[thbp]
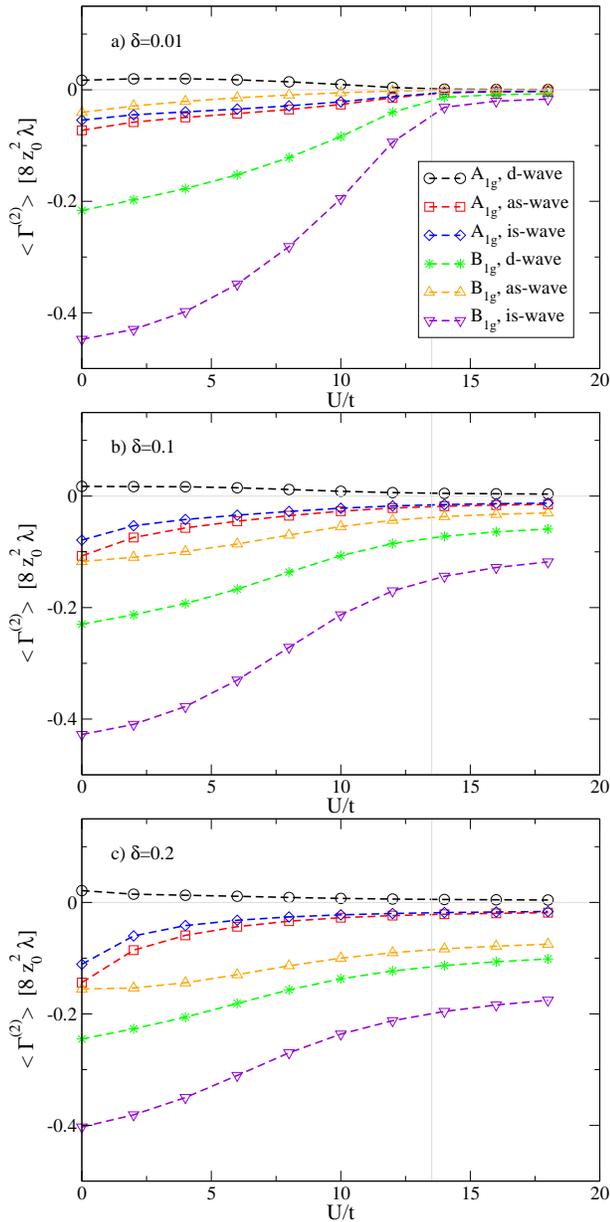

\includegraphics[width=8cm,clip=true]{fig9a.eps}
\includegraphics[width=8cm,clip=true]{fig9b.eps}
\includegraphics[width=8cm,clip=true]{fig9c.eps}
\caption{(Color online)
Average of the second order scattering function 
$\langle \Gamma_s^{(2)}(\kvec_F,\qvec)\rangle$ (cf. Eq. (\ref{eq:av})) 
over the Fermi surface
weighted with d-wave, anisotropic (as) and isotropic (is) 
s-wave symmetry functions.
$\delta=0.01$ (upper panel), $\delta=0.1$ central panel, $\delta=0.2$
lower panel; $t'/t=-0.4$. The vertical bar in the panels marks
the Brinkmann-Rice transition for the half-filled system.}
\label{fig:v1}
\end{figure}

In order to quantify the influence of correlations on superconductivity
we calculate the average of the scattering amplitude over the 
Fermi surface, weighted by the symmetry function for different
gap symmetries.
We consider the following function

\begin{equation}\label{eq:av}
\langle \Gamma \rangle = \frac{\frac{1}{N^2}\sum_{\kvec\pvec}
\delta(\epsilon_\kvec -E_F)\delta(\epsilon_\pvec -E_F)
\gamma_\kvec \gamma_\pvec \Gamma(\kvec,\pvec-\kvec)}
{\frac{1}{N}\sum_\kvec \delta(\epsilon_\kvec -E_F)(\gamma_\kvec)^2}
\end{equation}

and  $\gamma_\kvec$ specifies isotropic s-wave (is), anisotropic s-wave (as)
and d-wave (d) symmetry:
\begin{eqnarray*}
\gamma_\kvec&=&\left\lbrace
\begin{array}{lcl}
1 & &\mbox{(is)} \\
(\cos(k_x)+\cos(k_y))/2 & & \mbox{(as)}\\ 
(\cos(k_x)-\cos(k_y))/2 & & \mbox{(d)}.
\end{array}
\right.
\end{eqnarray*}

\begin{figure}[thbp]
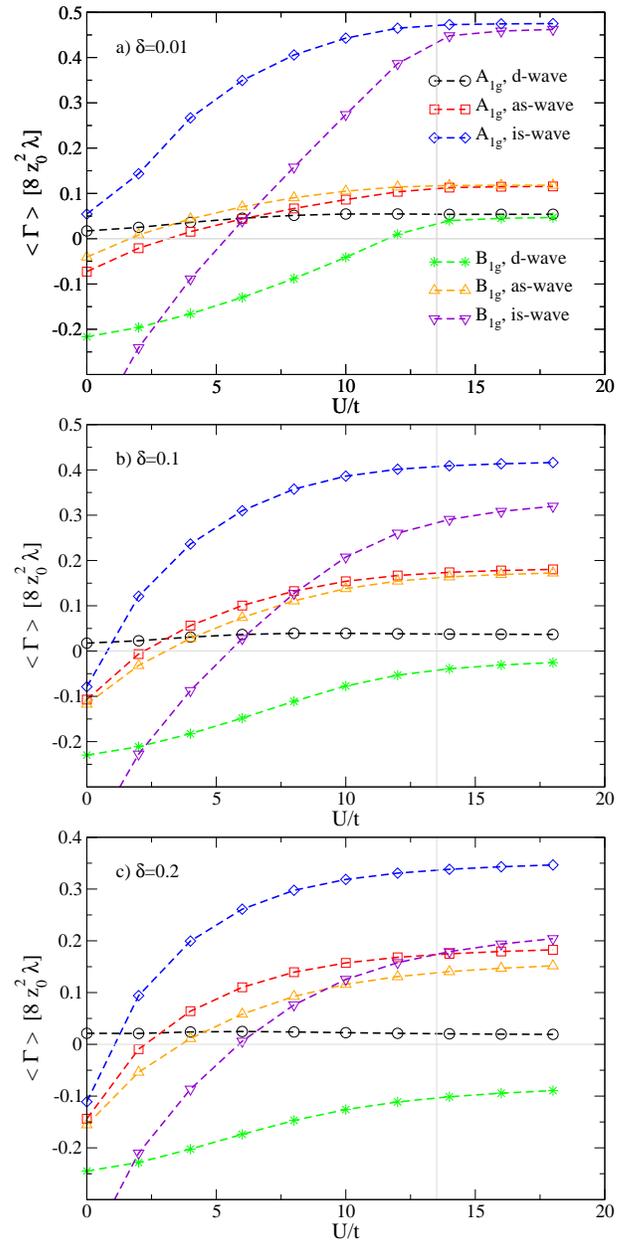

\includegraphics[width=8cm,clip=true]{fig10a.eps}
\includegraphics[width=8cm,clip=true]{fig10b.eps}
\includegraphics[width=8cm,clip=true]{fig10c.eps}
\caption{(Color online) Average of the full scattering function 
$\langle \Gamma_s(\kvec_F,\qvec)\rangle$ (cf. Eq. (\ref{eq:av})) 
over the Fermi surface
weighted with d-wave, anisotropic (as) and isotropic (is) 
s-wave symmetry functions.
$\delta=0.01$ (a), $\delta=0.1$ (b), and $\delta=0.2$ (c). 
$t'/t=-0.4$, $\lambda=g^2/\omega_0=0.5$. 
The vertical line in the panels marks
the Brinkmann-Rice transition for the half-filled system.}
\label{fig:v2}
\end{figure}

For the scattering amplitude $\Gamma(\kvec,\pvec-\kvec)$
we first use the second order expression Eq. (\ref{gamfin2nd}) in order
to study the influence of vertex corrections on SC separately (cf.
Fig. \ref{fig:v1}). Then the fully dressed expression Eq. (\ref{gamfin})
is used which, besides the repulsive Hubbard interaction 
Eq. (\ref{eq:eexp}) contains all higher order contributions to the
screening beyond ${\cal O}(g^2_{tr})$ (cf. Fig. \ref{fig:v2}). 

Fig. \ref{fig:v1} displays the average second order el-ph scattering amplitude 
for different dopings.  
Notice that a negative sign of $\langle \Gamma\rangle$
signals an attractive interaction whereas a positive value 
signals the suppression of the corresponding SC order 
(Ref. \onlinecite{johnston} uses the opposite convention).
The largest coupling is obtained for  B$_{1g}$ symmetry
where the isotropic s-wave pair scattering is more attractive than
that for d-wave pairs since we are analyzing only the renormalized
el-ph interaction without the direct effect of the Coulomb repulsion
on the scattering amplitude.  
In both channels, B$_{1g}$ and A$_{1g}$, and for all symmetries 
the average scattering  
from the  modes continuously decreases 
with increasing $U/t$ due to the reduction of the respective vertices. 
In the parameter range $U \ll U_{BR}$
it turns out that
$\langle \Gamma^{(2)}\rangle$ is not much affected by doping. 
On the other hand one finds that for 
$U/t$ beyond the Brinkmann-Rice limit
and doping close to half-filling (Fig. \ref{fig:v1}a) 
the scattering amplitudes becomes negligibly small but
start to increase for finite doping.

The average scattering from the A$_{1g}$ symmetric mode is small for both
s- and d-wave channels since $\langle \Gamma\rangle$
is only large when the form factor
of the SC pairs is small. Therefore as anticipated before the pairs
cannot take advantage from the correlation induced enhancement of the
vertex in this case.
For d-wave symmetry the effective el-el interaction even becomes
slightly repulsive. 
Notice that in the A$_{1g}$ channel the anisotropic s-wave scattering is slightly larger than that for isotropic s-wave pairs. However, upon analyzing
Eq. (\ref{eq:av}) we find that this effect is due to the normalization (i.e.
the denominator), whereas the numerator is larger for isotropic s-wave 
scattering as it should.

Finally we show in Fig. \ref{fig:v2} the result for the full
averaged pair scattering amplitude for both d- and s-wave
symmetries.
The inclusion of the Hubbard derived quasiparticle repulsion
Eq. (\ref{eq:eexp}) in the interaction part leads to further 
separation between d- and s-wave symmetries. Since this interaction is almost
structureless in momentum space it mainly leads to a
suppression of s-wave pair scattering (mostly pronounced for
isotropic s-wave) whereas the 
d-wave pairs are only weakly affected. The latter feel the local Coulomb
interaction mainly in the Mott-regime
where at half-filling the corresponding scattering now also becomes repulsive.
The inclusion of screening processes beyond ${\cal O}(g^2_{tr})$ 
via the RPA resummation starts to play
a role at higher doping where it reduces the strong dependence of
$\langle \Gamma\rangle$ on $U/t$.
According to our investigations the attractive scattering of d-wave pairs 
in the B$_{1g}$ channel dominates for $U/t \gtrsim 2.5$ over the corresponding 
isotropic s-wave pairing and for $U/t \gtrsim 5.5$ remains the only attractive
pairing symmetry within our model.

It should be mentioned that the GA correctly describes band narrowing and
the suppression of charge fluctuations but it fails to describe the
emergence of the magnetic $J$ scale in the metal as the 
Mott insulator is approached from
the Fermi liquid side. Therefore our present approach does not capture
pairing due to the exchange of magnetic fluctuations. Accordingly, if
we neglect the el-ph interaction the scattering amplitude becomes
repulsive in all the channels considered.

\section{Conclusion}\label{sec4}
The aim of the present paper was
to investigate the influence of an on-site electronic repulsion onto
a transitive electron-phonon coupling within a single-band Hubbard model.
In contrast to similar calculations for a local Holstein-type coupling
\cite{diciolo09} where the electron-phonon vertex is suppressed for all
momenta (but most strongly for large $\qvec$), we have seen that 
a transitive el-ph interactions leads to a much more 
interesting interplay between 
electronic correlations and phonons. Our finding that there can even be
a correlation-induced  enhancement of the pair scattering amplitude 
for certain momenta is especially important with regard to the
anisotropy of the el-ph interaction in cuprate superconductors \cite{dev04}.
In this regard our approach extends to the correlated regime, 
the analysis of Refs. \onlinecite{dev04,johnston},
which where based on a (electronically) non-interacting model.

The correlation-induced enhancement of the electron phonon interaction
it is not followed by a similar enhancement of the overall influence of
phonons on superconductivity. Even in the present case the impact of
phonons on superconductivity is suppressed by the Coulomb interactions. This is
because the enhancement happens in a channel that is not beneficial
for $d$-wave superconductivity, namely $A_{1g}$, and also because it
happens in a restricted momentum range. Notwithstanding the effect on
superconductivity, there may be a larger impact on other properties
which are more sensitive to the local, rather than the global, momentum
dependence like the appearance of kinks on the electron dispersion due
to self-energy effects. Also the enhancement of the scattering
function at small momentum can play an important role in phase
separation instabilities.\cite{cas95} 

We have presented our results as a function of $U$. 
A crucial question concerns  the strength of electronic correlations
in the high-T$_c$ materials. In fact, the early classification
of these compounds as doped Mott insulators has recently been questioned
in Ref.~\onlinecite{comanac}. This point of view agrees
with our own estimate for $U/t \approx 8$ as derived from a fitting
to the spin-wave dispersion of undoped lanthanum cuprates \cite{sei05,sei06}
and which is almost half the value of the Brinkman-Rice transition.

For $U/t=8$ we can deduce from Fig.~\ref{fig:v2} that the scattering
amplitude for d-wave scattering from the coupling to B$_{1g}$ modes
is reduced by $\sim 60\%$ for $\delta=0$, $\sim 50\%$ for $\delta=0.1$, 
and $40\%$ for $\delta=0.2$ which should be compared with the $20\%$
increase when the plasmon 
frequency is increased to $\Omega_{pl}=500 meV$.~\cite{johnston}
Thus even the inclusion of (moderate) correlation effects leaves
a significant contribution of phonons to d-wave superconductivity.

In case of the Hubbard-Holstein model \cite{diciolo09} the electron-phonon
vertex is determined by the local charge response function and it turned out
that the TDGA computation of this quantitiy is in rather good 
agreement with exact diagonalization 
and Quantum Monte Carlo results. 
Since for the present case of a transitive coupling detailed
computations from more sophisticated approaches are not yet 
available a detailed comparison concerning the performance
of the TDGA is not possible.
However, at least our previous investigations of charge
excitations on finite Hubbard clusters,~\cite{sei01,sei03} also including
the optical conductivity as an 'intersite' response function,
suggest that the TDGA is a reliable approach also for the issue
investigated in the present paper.
 
Our present study may also have relevance for an understanding of the
iron-based superconductors LiFeAs and NaFeAs which, contrary to the
other pnictide compounds, don't show magnetic order and have relatively
low $T_c$\cite{tapp08,parker08} which suggest that an electron-phonon
mechanism may
play  an important role. On the other hand, density functional
theory results \cite{jishi10} in coupling parameters which are far too small in 
order to account for the observed transition temperature. Recently, 
angle-resolved photoemission studies \cite{kordyuk10} have revealed both,
signatures of optical phonons and a significant electron-electron 
scattering rate in the spectra. One of the low energy phonon modes in 
the pnictides is an iron-arsenide bond-stretching phonon which 
affects the hopping between adjacent iron d-states in a manner similar
as the one analyzed here for the buckling modes.
Thus a natural perspective of the present approach would be its
combination with LDA computations for multiband Hubbard models
in order to obtain a realistic estimate of correlation effects
for the various phonon modes regarding their relevance for
the SC pairing in low $T_c$ pnictides. 

\acknowledgements
G.S. acknowledges financial support from the Deutsche Forschungsgemeinschaft.
J.L.'s research is partially supported by the Italian
Institute of Technology-Seed project NEWDFESCM.   
M.G. and J.L.  acknowledge  partial financial 
support from MIUR-PRIN 2007 (prot. 2007FW3MJX\_003).

\section*{Appendix} \label{apa}
The elements of the interaction kernel Eq. (\ref{Wqgen}) read as
\begin{eqnarray}                                                   
A_\qvec &=& V_\qvec -\frac{L_\qvec^2}{U_\qvec} \\                   
B_\qvec &=&                                 
\frac{1}{2} z_0 (z'+z_{+-}')- z_0 z'_D                                        
\frac{L_\qvec}{U_\qvec} \label{eq:bq}  \\                        
C_\qvec &=& - \ \  \frac{(z_0z_D')^2}{U_\qvec}                      
\end{eqnarray}  

with

\begin{eqnarray*}                                                                 
V_\qvec &=& \frac{e^0 z_0}{2}(z_{++}''+ 2z_{+-}''+z_{--}'') \\ 
&+&\frac{(z'+z_{+-}')^2}{2N}                                                 
\sum_{\kvec\sigma} \epsilon_{\kvec+\qvec,\sigma}^0n_{\kvec\sigma} \nonumber \\   
L_\qvec &=& e^0z_0 (z_{+D}''+ z_{-D}'')                                          
+\frac{z_D'(z'+z_{+-}')}{N}                                                      
\sum_{\kvec\sigma} \epsilon_{\kvec+\qvec,\sigma}^0n_{\kvec\sigma} \nonumber \\   
U_\qvec &=& 2e^0z_0 z_{D}''                                                      
+\frac{2(z_D')^2}{N}                                                             
\sum_{\kvec\sigma} \epsilon_{\kvec+\qvec,\sigma}^0 n_{\kvec\sigma} .
\end{eqnarray*}

Here we have defined the following abbreviations and derivatives of 
the hopping factors
\begin{eqnarray*}
 && z_{i\sigma}\equiv z_0 , \ \ \frac{\partial z_{i \sigma}}{\partial 
\rho_{ii \sigma}}\equiv z^{'},  \nonumber \\ &&  \frac{\partial z_{i\sigma}}
{\partial \rho_{ii -\sigma}}\equiv z^{'}_{+-},\               
\frac{\partial z_{i \sigma}}{\partial D_{i}}\equiv z^{'}_{D}                     
\label{z1} \\                                                                    
&& \frac{\partial^2 z_{i \sigma}}{\partial \rho^{2}_{ii \sigma}} \equiv 
z^{''}_{++}, \                                                                            
\frac{\partial^2 z_{i \sigma}}{\partial \rho_{ii \sigma}\partial \rho_{ii -\sigma}}                                                                            
\equiv z^{''}_{+-}, \ \frac{\partial^2 z_{i \sigma}}{\partial \rho^{2}_{ii -\sigma}}
 \equiv z^{''}_{--}  \nonumber \\                                             
&&  \frac{\partial^2 z_{i \sigma}}{\partial D^{2}_{i}}\equiv z^{''}_{D}, \ 
\frac{\partial^2 z_{i \sigma}}{\partial \rho_{ii \sigma}\partial D_{i}}\equiv z^{''}_{+D}, \ \frac{\partial^2 z_{i \sigma}}{\partial \rho_{ii -\sigma}\partial D_{i}}
\equiv z^{''}_{-D} 
\end{eqnarray*}                                                                  
Notice that in Ref. \onlinecite{diciolo09} 
the element $B_q$ is incorrect by a factor
of 'two'. Eq. (\ref{eq:bq}) reports the correct expression.


\begin{thebibliography}{9}
\bibitem{note1} The coupling to the local energy  
corresponds to a Holstein-type interaction which in connection with 
the Hubbard model has been discussed in Ref. \onlinecite{diciolo09}.
\end{thebibliography}
\end{document}